\documentclass[preprint,showpacs,preprintnumbers,amsmath,amssymb,authoryear]{revtex4}

\usepackage{ifthen}
\usepackage{amsmath} 
\usepackage[ruled,noend]{algorithm2e} 

\usepackage[english]{babel}
\usepackage{hyperref,pstricks,placeins,pst-text,multido}
\usepackage{pst-node, pst-plot}
\usepackage{todonotes}

\usepackage [autostyle, english = american]{csquotes}
\MakeOuterQuote{"}

\setcitestyle{authoryear,round}

\graphicspath{{./Fig/}}

\begin{document}

\title{A multi-GPU benchmark for 2D Marchenko Imaging }
\renewcommand{\thefootnote}{\fnsymbol{footnote}} 
\author{%
Victor Koehne, Matheu Santos, Rodrigo Santos, Diego Barrera, SENAI-CIMATEC HPC Center; Joeri Brackenhoff, ETH Zurich; and Jan Thorbecke, TU-Delft
}
\begin{abstract}
  The Marchenko method allows estimating Green's functions with a virtual source in the subsurface from a reflection response on the surface. It is an inverse problem that can be solved directly or by an iterative scheme, with the latter being more feasible computationally. In this work we present a multi-GPU implementation of a well-established iterative Marchenko algorithm based on (the) Neumann series. The time convolution and space integration performed on each iteration, also referred to as synthesis, are here represented as a segmented dot product, which can be accelerated on modern GPUs through the usage of warp-shuffle instructions and CUDA libraries. The original CPU version is benchmarked on 36 CPU cores versus the implemented version on 4 GPUs, over three different reflection data sets, with sizes ranging from 3 GB to 250 GB.
\end{abstract}
\maketitle

\section{Introduction}

The Marchenko method can estimate Green's function with a virtual source at an arbitrary focusing depth, using as inputs the reflection response measured at the surface and a background velocity model. The method and its iterative solution were introduced for 1D media by \cite{broggini2012connection} and later extended to 2D and 3D media by \cite{wapenaar2013three}. The implementation aspects of the iterative Marchenko method were first introduced by \cite{thorbecke2017implementation}, from which we can highlight: performing the convolutions in the frequency domain; transposition of the first two dimensions of the reflection response so that space integration is performed over the fastest dimension, i.e., the data layout is ordered as receivers, frequencies and shots; and estimating Green's functions for multiple focal points in parallel (e.g., with OpenMP). Regarding this last aspect the author reiterates that for one or few focal points, data loading becomes the main computational cost of the process. \cite{brackenhoff20203d} extends the aforementioned opensource code to three dimensions, and addresses the acceleration of data loading with the usage of compression and decompression via the ZFP library \citep{lindstrom2014fixed}.

\cite{ravasi2020implementation} address the computational issues of running a 3D Marchenko scheme on modern CPU clusters, provide open-source Python code, and discuss many implementation aspects. Although the data layout chosen, i.e., frequency slicing, for dividing the reflection response among computational resources is different from the one discussed here (referred to as slicing on row space, or receivers, frequencies, shots), the authors also conclude that the estimation of Green's functions for multiple focal points result in a performance advantage. \cite{broggini2018gpu} highlights the growth of graphics processing units (GPUs) on the composition of modern high performance computing clusters. As an accelerator, problems that best fit a GPU implementation are ones where the arithmetic intensity surpass the data transfers between CPU (host) and GPU (device) which is the case for the Marchenko scheme formulated as a series of segmented dot products. The author briefly discusses a GPU implementation of the Marchenko method, highlighting the gain in performance from estimating multiple Green's functions for many focal points in parallel, and the usage of optimized CUDA libraries, which are points we also discuss in this work.

\section*{Theory}

As shown in \cite{wapenaar2014marchenko} and \cite{thorbecke2017implementation} the method consists of iteratively estimating the focusing functions $f_{1}^{+}$ and $f_{1}^{-}$ to, at the end of the iterations, calculate the total Green's function $G$ (with its source redatumed to the desired subsurface focal point), as well as its downgoing $G^+$ and upgoing $G^-$ components. By making use of a practical abstraction based on Neumann series properties, the iterative process can be performed using only a single field $N_i$, from which the focusing and Green's functions can be calculated later.
Before performing the iterations, the fields are initialized as $N_{-1}(\mathbf{x}_R,\mathbf{x}_F,t) = f_{1,0}^{+}(\mathbf{x}_R,\mathbf{x}_F,t) = G_d(\mathbf{x}_R,\mathbf{x}_F,-t)$,
where the sub-index $-1$ indicates initialization, the sub-index $0$ indicates the first iteration, $\mathbf{x}_R$ denotes the positions of the receivers on the surface and $\mathbf{x}_F $ denotes the position of the desired focal point in the subsurface. The $G_d$ field is the first arrival seismogram with a source injected at position $\mathbf{x}_F$, which can be estimated with an eikonal solver or a finite-difference algorithm. The core equation of the iterative process is:
\begin{equation}
     N_i(\mathbf{x}_R,\mathbf{x}_F,-t) =
     \theta_{t} \int_{\partial \mathbb{D}_0}  \int_{t'} R(\mathbf{x}_R,\mathbf{x},t-t')\times N_{i-1}(\mathbf{x},\mathbf{x}_F,t')dt'd\mathbf{x}. \label{eq:syn1}
\end{equation}
Calculating $N_i$ through space integral and time convolution, as shown in equation \ref{eq:syn1}, is called synthesis \citep{thorbecke2017implementation}. It is the most expensive step of the method because it involves the integration of the entire reflection response $R$, which is the data from all shots acquired at the surface (after free-surface multiple elimination and deconvolution of the source wavelet). The internal integral in time is performed in the frequency domain, to speed up the process, so every iteration requires a direct and a reverse Fourier transform of field $N_{i}(\mathbf{x},\mathbf{x}_F,t)$. These transforms correspond to a very short time compared to the whole process, where the numerical integration is the most time consuming step. The inverse transform is necessary to apply the mask $\theta_{t}$ which eliminates events that occurred in $
|t|<t_d$, with $t_d$ being the first arrival traveltime from $\mathbf{x_F}$ to receivers, and maintain the validity of Marchenko's iterative equations \citep{wapenaar2014marchenko}. $f_1^+$ and $f_1^-$ and the auxiliary fields $f_2$ and $p^{-}$ can be updated along the iterative process by the accumulation of the proper $N_i$ terms, as detailed in \cite{thorbecke2017implementation}. The latter two fields can be used to determine $G$ \citep{wapenaar2013three}. It is also possible to calculate $G^+$ and $G^-$ as in \cite{wapenaar2014marchenko}, where estimating each component requires an extra synthesis. To perform imaging we use an equation from the double-focusing method as in \cite{staring2018source}, where fields $G^-$ and $f_1^+$ are time convolved and space integrated, resulting in $G^{-,+}(\mathbf{x}_F,\mathbf{x}_F,t)$, from which the local reflectivity at $\mathbf{x}_F$ can be obtained by taking its zero-time sample. Imaging can be performed on the CPU or the GPU, after the Marchenko iterations, and takes about $2\%$ of the total processing time; for this reason, this step is not explicitly addressed in the following sections.

\section{IMPLEMENTATION}

The (multi-)GPU version is written in CUDA C, and its pseudo-code is described in Algorithm 1, based on the CPU versions in \cite{thorbecke2017implementation} and \cite{brackenhoff20203d}. Algorithm 1 starts with the whole reflection data being read, Fourier transformed and stored on RAM. All first arrivals are read and copied to array $iRN$. The iterative process is then started, and the GPU version of synthesis is called (Algorithm 2). As the GPU memory is limited (the used Tesla V100 has 32 GB of global memory), synthesis must be done in a batched fashion. The first loop, instead of being on shots, is done on batches of shots. For each round of shot batches, a certain batch is transferred to each GPU ("copyH2D\_shot\_batch"). The terminology "H2D" is an acronym for "host to device" (i.e., CPU to GPU). $d\_Refl$ exists on each GPU and correspond to a different chunk of shots copied from $Refl$. The second loop, on the focal points, is now also done as rounds of batches. For each round of focal points, a chunk of $iRN$ is Fourier transformed and copied to each GPU ("FFT\_and\_copyH2D\_f0\_batch"). $d\_Fop$ is local to each GPU and contain the slices of the focal points for the current batch. Unlike $d\_Refl$, where each GPU has different chunks of shots, all $d\_Fop$ correspond to the same focal point data on each GPU. Figure \ref{fig:synthScheme} depicts the synthesis process on a batch of 3 shots and 2 focal points, where dot products between columns of slices of shots and $Ni$ are performed. A dot product is a pointwise multiplication  ("kernel\_mul\_batch"), followed by a summation, or reduction ("CUB\_segmented\_reduce"). In the case of many dot products of the same size, the problem can be understood as a segmented dot product (segmented reduction), where the segment size is the column size ($nr$ receivers, in this case). 

Since the release of the NVIDIA Kepler architecture in 2014, a low level instruction denominated warp shuffle is available and can be used to accelerate reductions on the GPU. This instruction (SHFL) enables a thread to directly read a register from another thread in the same warp, and has many advantages over shared memory, namely: shuffle replaces a multi-instruction shared memory sequence with a single instruction, increasing effective bandwidth and decreasing latency;
shuffle does not use any shared memory;
synchronization is within a warp and is implicit in the instruction, so there is no need to synchronize the whole thread block \citep{luitjens2014faster}. Many optimized CUDA libraries can perform reduction, normal and/or segmented, such as Thrust \citep{bell2012thrust} and CUB \citep{merrill2015cub}, with CUB giving the best results, and for this reason, being the chosen library in this work. It is important to highlight that CUB does not provide a segmented dot product function (only segmented reduction), and for this reason it is necessary to implement a pointwise multiplication kernel before calling CUB. The batched synthesis is finished by padding, inverse Fourier transforming the corresponding $d\_sum$ from each GPU to the corresponding traces of the host $iRN$, which is then saved to disk. The loops continue with different batches until all data has been processed. All $iRN$ are then time-reversed, multiplied by $-1$ and muted, and will be batched to the GPUs on the next iteration. While the CPU version calculates arrays $Ni$, $f2p$, $f1plus$, $f1min$ and $pmin$ on the fly, the GPU version does it afterwards (on the CPU) by reading previously saved $iRN$ from disk. This is done to save memory (and has an extra read/write cost discussed later), as only the $iRN$ array must be batched to the GPUs. With $G^-$ and $f_1^+$, the image amplitudes for each focal point are then estimated.

\section{RESULTS}

\begin{figure}[h!]
\centering
    \includegraphics[width=1.0\textwidth]{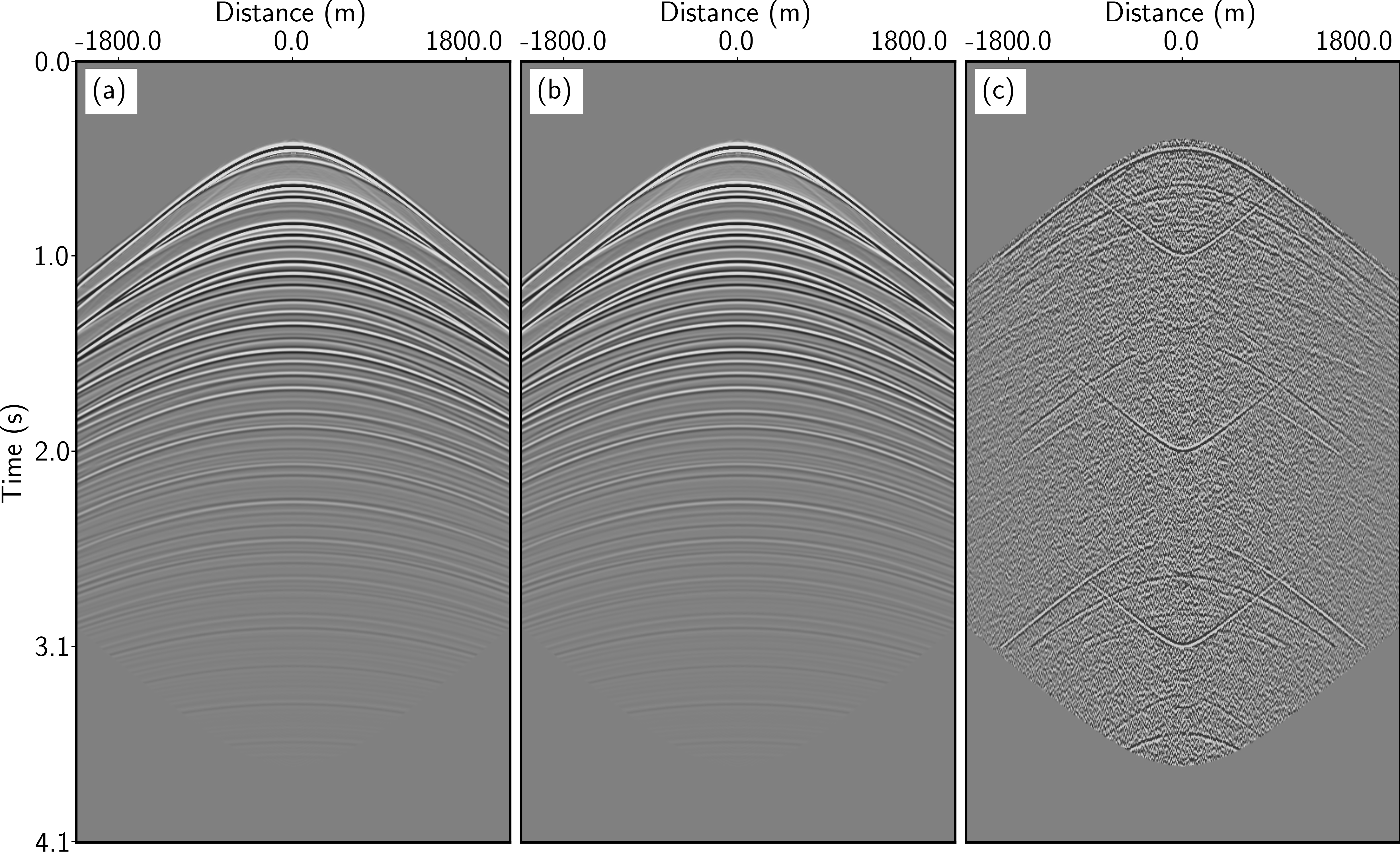}
    \vspace{-0.7cm}  
    \caption{Retrieved Green's functions of $\mathbf{x_F}=(0,900)$ m. (a) CPU version; (b) GPU version; (c) Difference (NRMSE=$10^{-6}$).}
    \label{fig:green}
\end{figure}

\begin{figure}[h!]
\centering
    \includegraphics[width=1.0\textwidth]{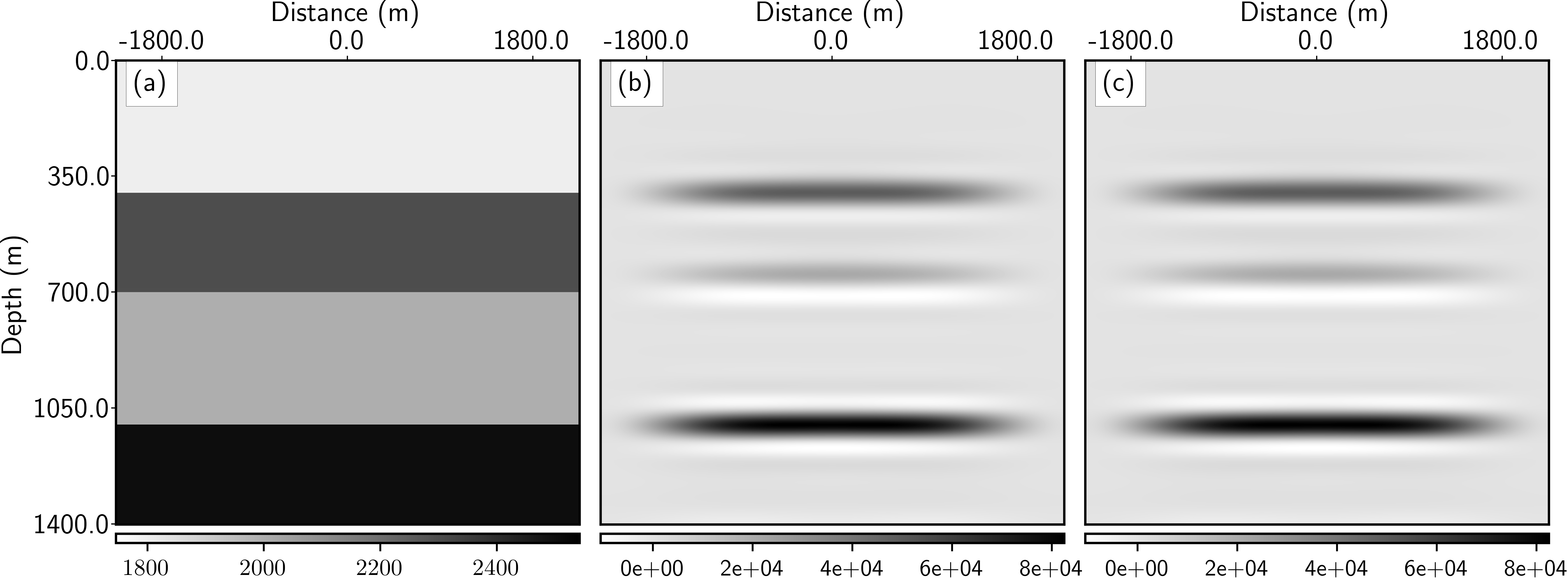}
    \vspace{-0.7cm}  
    \caption{(a) Velocity model for 901 shots example. (b) CPU image. (c) GPU image. Difference (not shown) has NRMSE=$10^{-5}$. Gain of $t^3$ applied for plotting purposes.}
    \label{fig:images}
\end{figure}

We perform our tests in a single node of a supercomputer with the following configurations: 376 GB of RAM (DDR4) memory, 36 Intel(R) Xeon(R) Gold 6240 CPU @ 2.60GHz processors and 4 NVIDIA(R) Tesla V-100 with 32 GB of global memory each. In this manner, the CPU version is benchmarked using 36 cores (parallelized with OpenMP), while the multi-GPU version uses 4 V100 for the synthesis, and the same 36 cores for the auxiliary processes. This performance benchmark is in line with disclosed Tesla V100 benchmarks \citep{nvidia2018tesla}, but single-core speedups are also presented. We measure the performance of CPU and multi-GPU versions on three different versions of the flat layers model (Figure \ref{fig:images}-(a)), or \textit{oneD} as in \cite{thorbecke2017implementation}: (a) with a depth of 1.4 km and an extension of 10km, 901 sources/receivers and a disk size of 3.3 GB for $R$; (b) with a depth of 1.4 km, an extension of 20 km, 1901 sources/receivers and a disk size of 30 GB for $R$; (c) with a depth of 1.4 km, an extension of 40km, 3901 sources/receivers and a disk size of 249 GB for $R$. In all three models, Marchenko imaging with 8 iterations was tested for 1, 29 and 145 focal points. Figure \ref{fig:green} shows retrieved Green's functions of a single focal point on the CPU (\ref{fig:green}-(a)) and GPU (\ref{fig:green}-(b)), and their difference (\ref{fig:green}-(c)), which shows a normalized root mean squared error of $10^{-6}$. Similarly, Figures \ref{fig:images}-(b,c) shows retrieved images for 901 shots and 145 focal points for the CPU and GPU versions, respectively, with a NRMSE of $10^{-5}$. These differences indicate that the GPU version is accurate to within certain amount above floating-point precision.

\begin{figure}[h!]
\centering
     \includegraphics[scale=0.65]{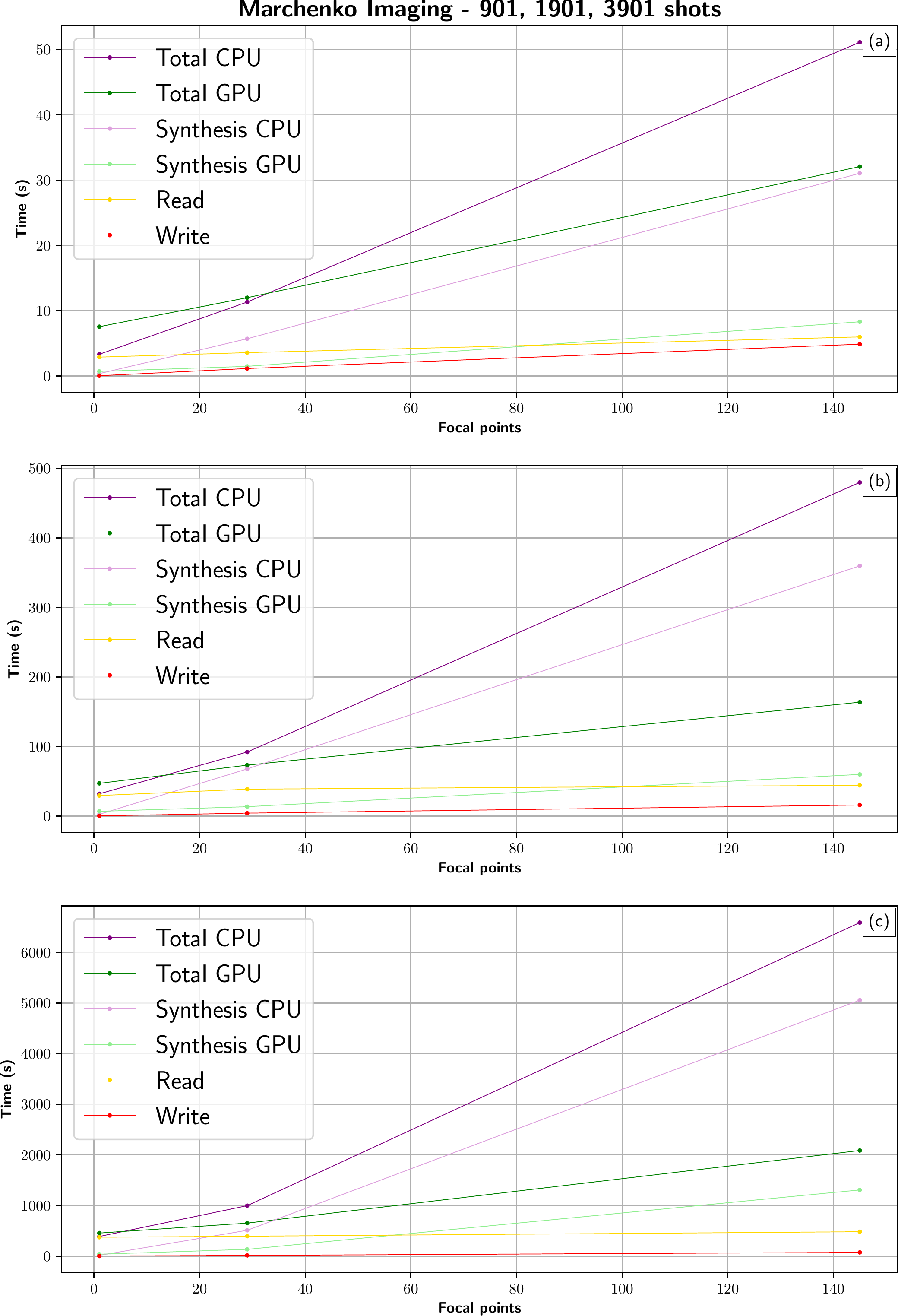} 
    \caption{Marchenko imaging times for CPU and GPU versions, for data sets of: 901 shots (3.3 GB), 1901 shots (30 GB) and 3901 shots (249 GB). Each set of tests were run for 1, 29 and 145 focal points.}
    \label{fig:times}
\end{figure}

\begin{figure}[h!]
\centering
\begingroup
 \setlength{\tabcolsep}{0.025cm} 
  \renewcommand{\arraystretch}{0.8} 
    \begin{tabular}{cc}
         \includegraphics[width=0.4\textwidth]{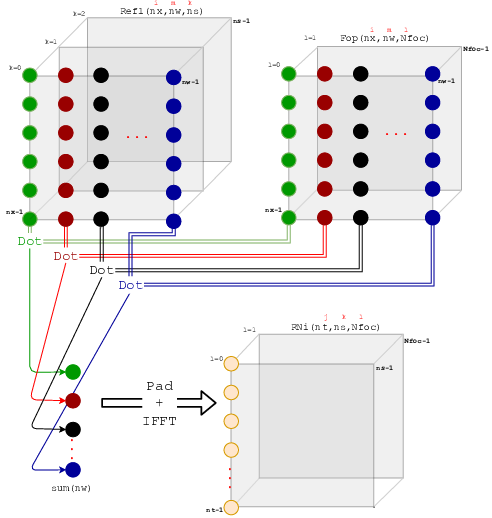} & \includegraphics[width=0.4\textwidth]{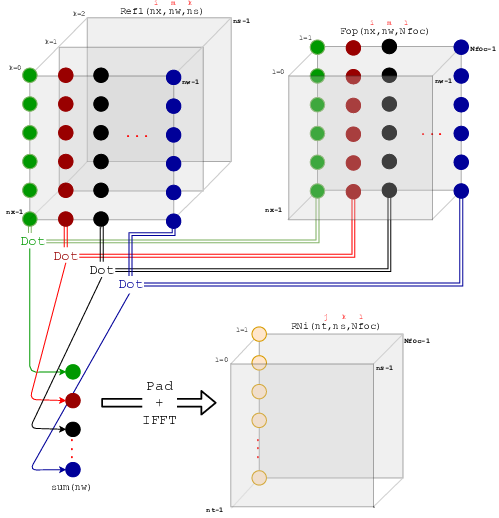} \\ \includegraphics[width=0.4\textwidth]{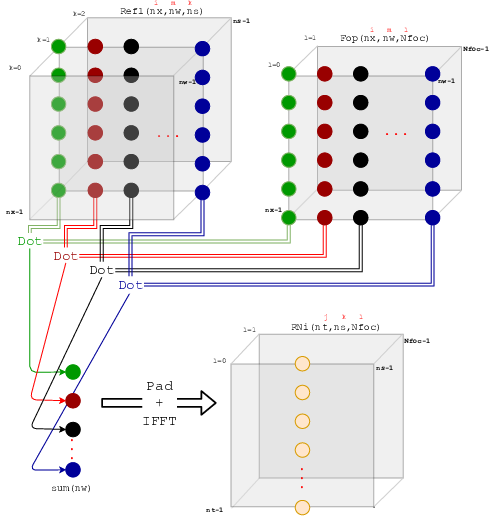} &
         \includegraphics[width=0.4\textwidth]{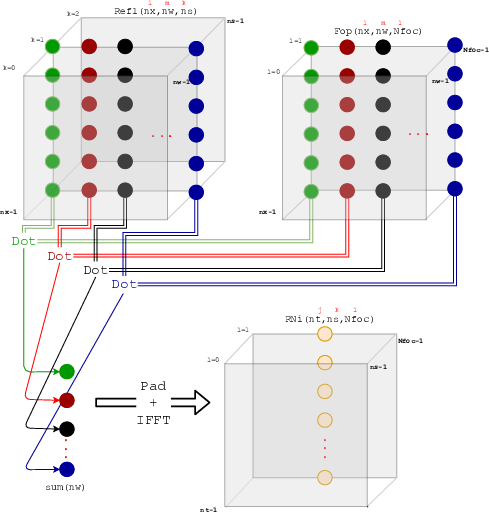} \\ \includegraphics[width=0.4\textwidth]{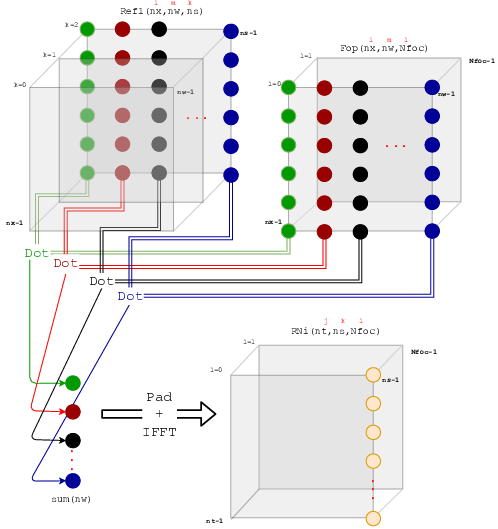} & \includegraphics[width=0.4\textwidth]{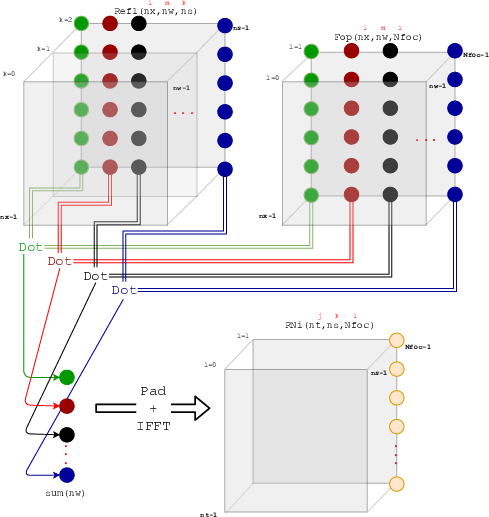} 
    \end{tabular}
\endgroup
    \vspace{-0.4cm}  
    \caption{Synthesis scheme for a batch of 3 shots (k=0,1,2) and 2 first arrivals (l=0,1). Each trace of $N_i$ depend on a single k and l. \label{fig:synthScheme}}
\end{figure}

Figure \ref{fig:times}-(a) shows the Marchenko scheme times for imaging of 1, 29 and 145 focal points with 901 shots and receivers. We start by evaluating only the synthesis times, i.e., the light purple (36 CPU cores) and light green curves (4 GPUs). The speedups of the multi-GPU over the multi-core versions are: 0.51, 3.85 and 3.73 for 1, 29 and 145 focal points, respectively. Comparing to a single-core run, the groups of speedups are 8.78, 15.87, 14.42 for the multi-core version and 4.51, 61.15, 54.82 for the multi-GPU version. The speedups show that the multi-GPU version is better only if a certain number of imaged focal points are batched together; comparing the full-time curves (dark purple and dark green), this number is around 35 focal points. As shown in Figure \ref{fig:synthScheme}, the same batch of shots goes through a segmented dot product with different $Fop$ (focal point) matrices to estimate different traces of the final output $RNi$; in other words, estimating a single focal point has less data re-usage than estimating many focal points. As the GPU synthesis times include data transfers between host and device, the synthesis FLOPs per bytes transferred ratio shows a performance gain only after 35 estimated focal points, and decline for 149 focal points. Finding the best number of focal points that would give the most speedup depends on many factors (hardware, workload size) and could be explored in future works. Still on Figure \ref{fig:times}-(a), we see a yellow and a red curve which correspond to read and write times. The read curve refers to both the CPU and GPU implementations and depicts reading of the reflection response and first arrivals from the disk. For the CPU version, these times correspond to 87\%, 32\% and 11\% of the total time for 1, 29, 145 focal points; and 37\%, 29\% and 18\% for the GPU version. This shows that for very few focal points data loading becomes the main bottleneck; \cite{brackenhoff20203d} address ways of accelerating this step on Marchenko by focusing on data compression. The red (write) curve corresponds only to the GPU version and refers to the modification of the original Marchenko code (writing $N_i$ at every iteration) to save GPU memory. These times correspond to 0.5\%, 9.5
\% and 15\% of total times; though the writing is not a big bottleneck, the overall code performance could improve by writing in parallel with a dedicated CPU thread.

As it can be seen from Figures \ref{fig:times}-(b,c), the same results described previously are maintained for bigger data sets. Figures \ref{fig:times}-(b) shows times for 1901 shots (12 GB). The synthesis speedups for GPU vs CPU versions are 0.34, 5.09 and 6.02 (4.17, 65.77, 307.02 compared to 1 core) for 1, 29 and 145 focal points. The total speedups are 0.68, 1.25, 2.93 (1.54, 10.98, 19.81 compared to 1 core). The number of batched focal points that makes the multi-GPU version more advantageous is 15. Figure \ref{fig:times}-(c) shows times for 3901 shots (249 GB). The synthesis speedups for GPU vs CPU versions are 0.39, 3.83 and 3.87 (7.83, 60.47, 85.15 compared to 1 core) for 1, 29 and 145 focal points. The total speedups are 0.63, 1.52, 3.16 (1.76, 14.60, 58.47 compared to 1 core). The number of batched focal points that makes the multi-GPU version more advantageous is now around 3. This can be considered the most important example since in the other two the whole reflection response and the $N_i$ cubes fit the GPU global memory (36 GB), while this test forces pipelining of both reflection data and the $N_i$ cubes. The running times maintain their trend over the CPU version, showcasing that pipelining data on to GPUs is feasible for the 3D case where it is almost always necessary. The 3D problem is highly parallelizable and can be interpreted as slices on Figure \ref{fig:synthScheme} becoming cubes. The presented expanded abstract, together with \cite{brackenhoff20203d} 3D extension of the 2D CPU scheme can serve as a basis for future 3D Marchenko multi-GPU implementations.

\section{CONCLUSIONS}

We presented a multi-GPU adaptation of a well-established multi-core code for iterative Marchenko imaging. The main kernel of the scheme is formulated as a segmented dot product and accelerated with warp-shuffle instructions through the CUDA CUB library. Previous literature shows that the method has better performance if many focal points are imaged in batch, which is confirmed for this multi-GPU version. The results are presented for data sets of different sizes ranging from 3 to 250 GB, showing a similar speedup trend. The 250 GB example showcases an efficient pipeline of reflection and first arrival data to GPUs and paves the way for a future 3D multi-GPU implementation of the Marchenko scheme.  

\section{ACKNOWLEDGMENTS}

This research was performed in the framework of the project "Marchenko CENPES" financially supported by Centro de Pesquisas Leopoldo Americo Miguez de Mello (CENPES) at Petrobras. We wish to thank CENPES and SENAI CIMATEC for permission to publish this research, carried out at the SENAI CIMATEC Supercomputing Center (CS2I). We also acknowledge R. Pestana for the comments and suggestions which greatly improved the text.

{\SetAlgoNoLine 
\begin{algorithm}
\DontPrintSemicolon 
Main \textbf{begin}\;
\quad Reading SU-style input Data and Allocate arrays\;
\quad CPU Initialization\;
\quad GPU(s) Initialization\;
\quad Read all shots, all first arrivals to RAM\;
\quad iRN(t) = G\_d(-t)\;
\quad \For{iter$\leftarrow$0 \KwTo niter}
    { 
\quad \quad cuda\_synthesis(Refl, iRN, iRN)\;
\quad \quad write(iRN, "iRN\_iter")\; 
\quad \quad iRN(t) = -iRN(-t)\;
\quad \quad applyMute(iRN, muteW)\;
\quad \textbf{end}\; 
    }
\vspace{0.25cm}
\quad Ni(t) = f2p(t) = f1plus(t) = G\_d(-t)\;
\quad f1min(t) = pmin(t) = 0.0 \;
\quad \For{iter$\leftarrow$0 \KwTo niter}
    {
\quad \quad read(iRN, "iRN\_iter")\;
\quad \quad Ni(t) = -iRN(-t)\;
\quad \quad pmin(t) += iRN(t)\;
\quad \quad applyMute(Ni, muteW)\;
\quad \quad f2p(t) += Ni(t)\;
\quad \quad \uIf{ (iter\%2==0) }
            {
\quad \quad \quad f1min(t) -= Ni(-t)\;
            }
\quad \quad \uElse
            {
\quad \quad \quad f1plus(t) += Ni(t)\;
            }
\quad \textbf{end}\; 

\quad cuda\_synthesis(Refl, f1plus, iRN)\; 
\quad Gmin(t) = iRN(t) - fmin(t)\;    
\quad imaging(Gmin, f1plus, image)\;
\textbf{end}\; 
    }

\caption{GPU implementation of the iterative Marchenko algorithm.}
\end{algorithm}
}

{\SetAlgoNoLine 
\begin{algorithm}[h!]
\DontPrintSemicolon 
cuda\_synthesis(Refl, iRN)\;
\textbf{begin}\;

\quad \For{isb$\leftarrow$0 \KwTo Nbatch\_shots} 
    {
\quad \quad \ForEach{GPU}{ 
\quad \quad \quad copyH2D\_shot\_batch(Refl, d\_Refl)\;}
\quad \quad \textbf{end}\; 
\quad \quad \For{isf$\leftarrow$0 \KwTo Nbatch\_f0s} 
            {
\quad \quad \quad \ForEach{GPU}{ 
\quad \quad \quad \quad FFT\_and\_copyH2D\_f0\_batch(iRN, d\_Fop)\;}
\quad \quad \quad \textbf{end}\; 
\quad \quad \quad \ForEach{GPU}
                { 
\quad \quad \quad \quad kernel\_mul\_batch(d\_Refl, d\_Fop, d\_RF)\;
\quad \quad \quad \quad CUB\_segmented\_reduce(d\_RF, d\_sum)\;
\quad \quad \quad \quad IFFT\_and\_copyD2H(d\_sum, iRN)\;
\quad \quad \quad \textbf{end}\; 
                }
\quad \quad \textbf{end}\; 
            } 
\quad \textbf{end}\; 
    }
\textbf{end}\; 
\caption{CUDA synthesis.}
\end{algorithm}
}

\vspace{2cm}

\clearpage

\bibliographystyle{seg}  
\bibliography{example}

\end{document}